\documentclass[letterpaper, 10 pt, conference]{ieeeconf}  

\IEEEoverridecommandlockouts                              

\usepackage{graphics}

\usepackage{amsmath}
\usepackage{amssymb}
\usepackage{amsfonts}
\usepackage{mathrsfs}	
\usepackage{bm}
\usepackage{mathtools} 
\usepackage[dvipsnames]{xcolor}
\usepackage[colorlinks=true,allcolors=black]{hyperref}
\usepackage{tikz}
\usetikzlibrary{positioning,shapes,arrows.meta,bending}

\newtheorem{lemma}{Lemma}
\newtheorem{theorem}{Theorem}
\newtheorem{definition}{Definition}
\newtheorem{assumption}{Assumption}
\newtheorem{proposition}{Proposition}
\newtheorem{remark}{Remark}

\newcommand{\eye}{\mathbb{I}}
\newcommand{\tp}{^\mathsf{T}}

\newcommand{\cpem}[1]{{S}(#1)}

\renewcommand{\QED}{\QEDopen}

\title{\LARGE \bf A Symmetry-Preserving Reduced-Order Observer}

\author{Jeremy W. Hopwood$^{1}$ and Craig A. Woolsey$^{2}$
\thanks{*This work was supported by NASA under Grant No.~80NSSC20M0162.}
\thanks{$^{1}$Jeremy W. Hopwood is a Ph.D. Candidate in the Kevin T. Crofton Department of Aerospace and Ocean Engineering, Virginia Tech, Blacksburg, VA 24061, USA
        {\tt\small jeremyhopwood@vt.edu}}%
\thanks{$^{2}$Craig A. Woolsey is a Professor in the Kevin T. Crofton Department of Aerospace and Ocean Engineering, Virginia Tech, Blacksburg, VA 24061, USA
        {\tt\small cwoolsey@vt.edu}}%
}

\begin{document}

\maketitle
\thispagestyle{empty}
\pagestyle{empty}

\begin{abstract}
A symmetry-preserving, reduced-order state observer is presented for the unmeasured part of a system's state, where the nonlinear system dynamics exhibit symmetry under the action of a Lie group. Leveraging this symmetry with a moving frame, the observer dynamics are constructed such that they are invariant under the Lie group's action. Sufficient conditions for the observer to be asymptotically stable are developed by studying the stability of an invariant error system. As an illustrative example, the observer is applied to the problem of rigid-body velocity estimation, which demonstrates how exploiting the symmetry of the system can simplify the stabilization of the estimation error dynamics. 
\end{abstract}

\section{INTRODUCTION}

Methods for designing state observers for nonlinear systems are limited, and there are no general techniques that guarantee global convergence of the estimation error as there are in the linear case~\cite[Ch.~15]{rughLinearSystemTheory1996}. Provably effective state estimation strategies are inevitably limited to special classes of systems, motivating considerable attention in the past several decades to nonlinear observers. Early results generally rely either on Lipschitz conditions~(e.g.,~\cite{thauObservingStateNonlinear1973,gauthierSimpleObserverNonlinear1992,rajamaniObserversLipschitzNonlinear1998}) or finding a transformation of the system to some canonical form~(e.g.,~\cite{krenerNonlinearObserversLinearizable1985,phelpsConstructingNonlinearObservers1991,yixiongSlidingModeObserver2001}). For some systems, however, these methods are overly-conservative or difficult to implement. Passivity-based observers~(\cite{shimNonlinearObserverDesign2003,venkatramanFullorderObserverDesign2010}) constitute an approach that aims to overcome these deficiencies by viewing observer design as output feedback control of a state estimation error system --- a perspective inherently shared among nonlinear observer techniques as a result of the duality between nonlinear controllability and observability~\cite[Ch.~3]{nijmeijerNonlinearDynamicalControl1990}.

Approaches to observer design that leverage the structure of the system are of particular interest. Specifically, the role of differential geometry in observer design has been explored, in which symmetries of a nonlinear dynamical system are preserved in the state observer and its state estimation error dynamics, aiding in design and stability analysis. From a Lagrangian perspective, Aghannan and Rouchon~\cite{aghannanIntrinsicObserverClass2003} leveraged the symmetry inherent in the coordinate-free Euler-Lagrange equations to design a nonlinear observer given measurements of the system's configuration. This idea of leveraging symmetries in the dynamics was generalized by Bonnabel et~al. to include nonlinear systems under a Lie group's action~\cite{bonnabelSymmetryPreservingObservers2008} as well as systems defined on Lie groups~\cite{bonnabelNonLinearSymmetryPreservingObservers2009}. The idea of these so-called \emph{symmetry-preserving observers} is to design an observer that is also invariant, i.e., for which the observer dynamics also preserve this Lie group symmetry. This approach allows the observer's convergence properties to be analyzed more easily because of simplifications afforded by the system symmetry. The invariant extended Kalman filter (EKF) developed in Bonnabel et~al.~\cite{bonnabelInvariantExtendedKalman2009} is a constructive, local symmetry-preserving observer that is also applicable to systems with measurement and process noise. The invariant EKF was shown by Barrau and Bonnabel in~\cite{barrauInvariantExtendedKalman2017} to be a stable observer, admitting an exact linearization for group affine systems. From a preservative of \emph{equivariance}, Mahony et~al.~\cite{mahonyObserversKinematicSystems2013} developed a nonlinear observer for kinematic systems with complete symmetry --- that is, systems defined on homogeneous spaces. This work was generalized with the equivariant Kalman filter, which applies to general equivariant systems (see van Goor et~al.~\cite{vangoorEquivariantFilterEqF2023} and references therein). 

Existing approaches to symmetry-preserving observers only consider the \emph{full-order} case, however, in which the entire state of the system is estimated. In many scenarios, part of the system's state may be known with negligible error or may be obtained as the output of an observer whose design is independent of the rest of the system's state. For example, attitude observers for aircraft or spacecraft often do not rely on the rigid body's translational dynamics~(e.g., \cite{leffertsKalmanFilteringSpacecraft1982} and \cite{mahonyNonlinearComplementaryFilters2008}). Another example is the problem of wind estimation from aircraft motion~(e.g., \cite{gonzalez-rochaSensingWindQuadrotor2019,chenInvariantEKFDesignQuadcopter2022}), where the main goal is to obtain estimates of wind and air-relative velocity --- not to re-estimate the aircraft's position, attitude, and angular velocity. Estimating wind can be framed as a disturbance estimation problem, falling into the general category where the internal state of the system is known but the disturbance is not. Disturbance observers generally assume the dynamics of the disturbance are unknown~(e.g., \cite{khalilExtendedHighGainObservers2017,chenStructureInspiredDisturbanceObserver2024}) and can also be applied to systems defined on Lie groups (e.g., \cite{shiExtendedStateObserverBasedFiniteTime2021,wangGOBGeometricObserver2024}). However, when the disturbance dynamics are known, they may be leveraged to obtain improved performance and/or stronger guarantees.

In scenarios where full-order observers are either unnecessary or impractical, \emph{reduced-order observers}, in the sense of Karagiannis~et~al. in \cite{karagiannisNonlinearObserverDesign2005}~and~\cite{astolfiChapter5Reducedorder2008}, are of particular interest in which only the unmeasured part of the system's state is estimated. The aim of reduced-order observer design is to render a particular set, characterized by zero state estimation error, positively invariant and globally asymptotically attractive. However, the definitions of this set and the observer dynamics are non-constructive (although constructive approaches are available for certain classes of systems~\cite{venkatramanFullorderObserverDesign2010}). In this paper, we develop a reduced-order observer that is also symmetry-preserving. That is, the zero-error set and the observer dynamics are constructed in a way such that they are invariant under the Lie group's action. Leveraging this symmetry leads to an invariant state estimation error system, simplifying observer parameter selection and stability analysis.

The remainder of this paper is organized as follows. Section~\ref{sec:preliminaries} introduces the preliminary concepts that will be used in the development of the pre-observer in Section~\ref{sec:pre-observer}. Next, sufficient conditions for the pre-observer to be an asymptotically stable observer are presented in Section~\ref{sec:observer}. Finally, the main results are applied to the example of rigid-body velocity estimation in Section~\ref{sec:example}, followed by concluding remarks in Section~\ref{sec:conclusions}.

\section{PRELIMINARIES}\label{sec:preliminaries}

\subsection{Transformation Groups, the Moving Frame, and Invariant Dynamics}

\subsubsection{Transformation Groups~\cite{boothbyIntroductionDifferentiableManifolds2003,olverClassicalInvariantTheory1999}}
Consider a differentiable (i.e., $C^\infty$ or smooth) manifold~$\mathcal{X}$ on which a Lie group $G$ acts via the mapping
\begin{equation*}
	\varphi: G \times \mathcal{X} \to \mathcal{X},~(g,x) \mapsto \varphi_g(x)
\end{equation*}
such that (i)~the identity element $e$ in $G$ induces the identity transformation $\varphi_e(x) = x$ for all $x\in\mathcal{X}$, and (ii)~ the composition of group actions satisfies $\varphi_g \circ \varphi_h = \varphi_{g \ast h}$, where ``$\circ$'' denotes the composition of mappings and ``$\ast$'' is group multiplication. The inverse transformation $\varphi_g^{-1}$ is given by the action of the inverse group element -- i.e., $\varphi_g^{-1} = \varphi_{g^{-1}}$. The Lie group $G$ is said to \emph{act freely} on $\mathcal{X}$ if $\varphi_g(x) = x$ implies $g$ is the identity element, $e$. The collection $\{\varphi_g\}_{g\in G}$ is called a \emph{transformation group}. The \emph{$G$-orbit} of a point $x\in\mathcal{X}$ is the set $\{ \varphi_g(x) ~|~ g \in G\}$. 

\subsubsection{The Moving Frame~\cite{bonnabelSymmetryPreservingObservers2008,olverClassicalInvariantTheory1999,mansfieldModernNotionMoving2011}}
\begin{figure}[htb!]
	\centering
	\vspace{0.5em}
	\begin{tikzpicture}[x=1em,y=1em]
		\tikzstyle{every node}=[font=\small]
		\definecolor{light_blue}{HTML}{56b4e9}
		%
		\node (X) at (0,0) {\includegraphics[width=20em]{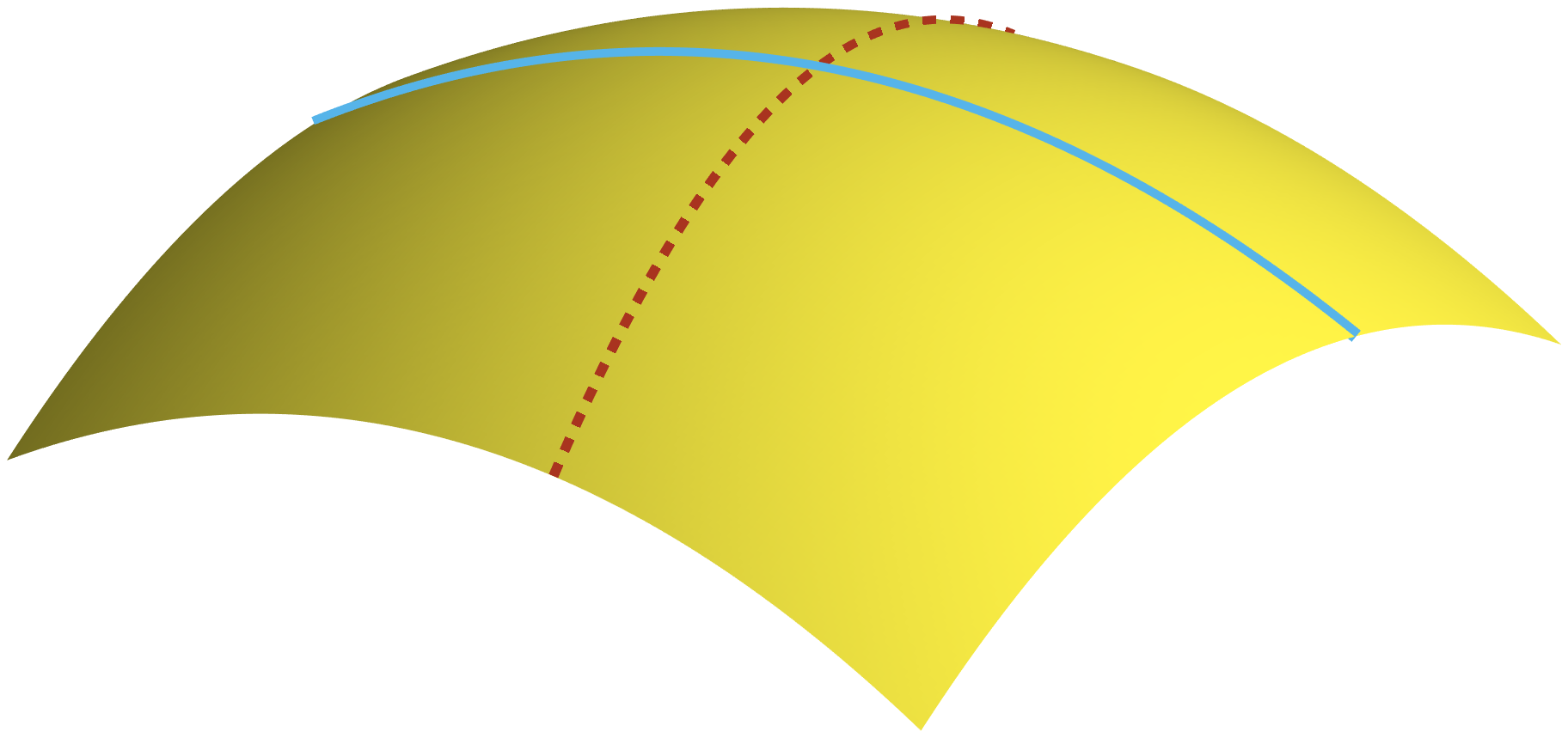}};
		\node (G) at (8,-7) {\includegraphics[width=14em]{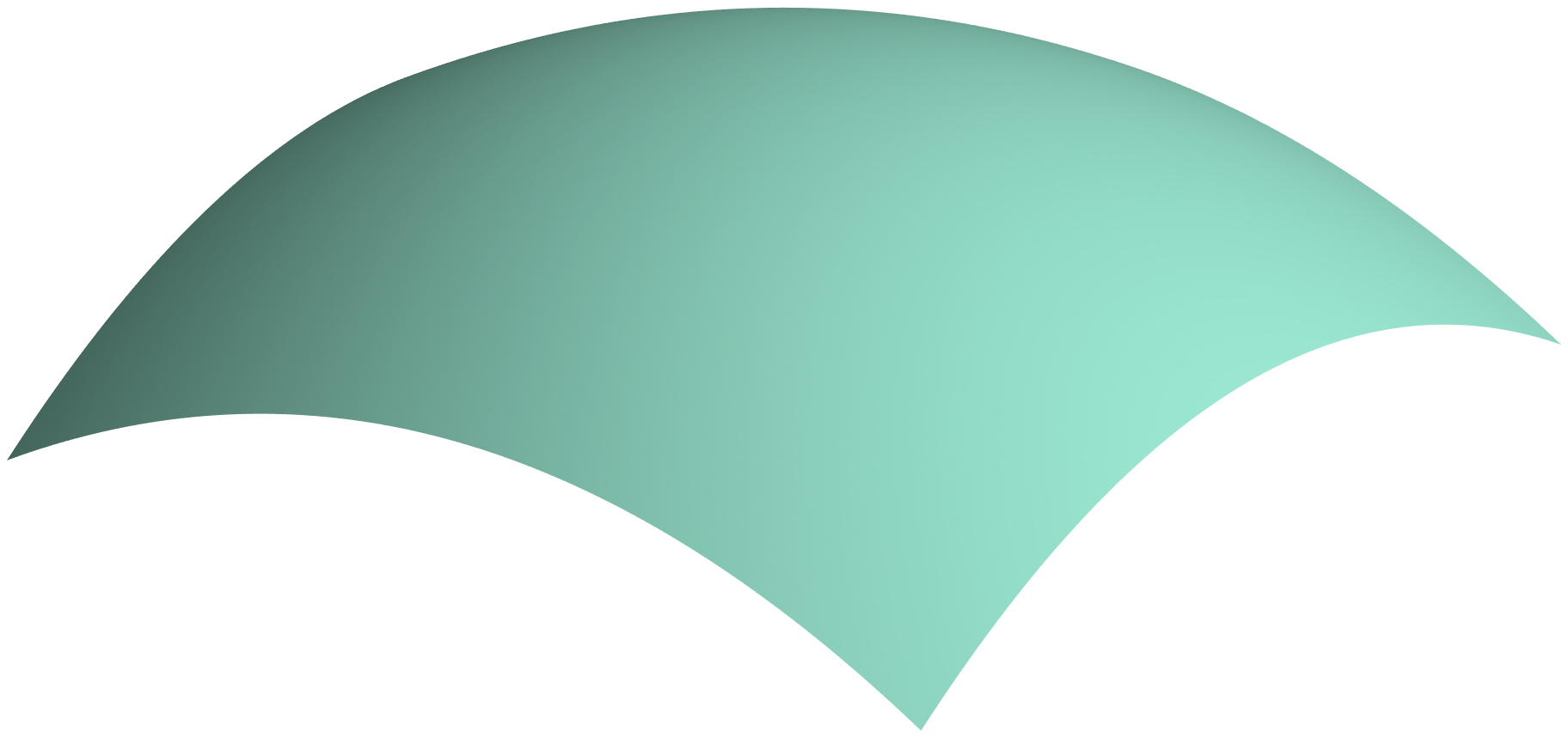}};
		%
		\coordinate[label=right:$x$, above left = 2 and 0.9 of X.center] (x);
		\coordinate[label=right:$\varphi_g(x)$, below left = 1.8 and 1.1 of x] (phi);
		\coordinate[above right = 1.5 and 1.3 of x] (phi_gamma);
		\coordinate[label=left:$\gamma(x)$, above left = 1.5 and 1.5 of G.center] (gamma);
		\coordinate[label=left:$\gamma(\varphi_g(x))$, below right = 0.4 and 2 of G.center] (gamma_phi);
		\fill[black] (x) circle (2pt);
		\fill[black] (phi) circle (2pt);
		\fill[black] (gamma) circle (2pt);
		\fill[black] (gamma_phi) circle (2pt);
		%
		\draw[light_blue,line width=1.9pt] (phi_gamma) circle (3pt);
		\node[rectangle callout,fill=white,rounded corners,draw=light_blue,line width=1pt,callout absolute pointer=(phi_gamma.north),above left= 1em and 1em of phi_gamma] {$\varphi_{\gamma(x)}^{\mathrm{inv}}(x)=k$};
		\node[rectangle callout,fill=white,rounded corners,draw=black,line width=1pt,callout absolute pointer=(phi_gamma.center),above right= 1em and 1em of phi_gamma] {$\varphi_{\gamma(x)}(x)$};
		\fill[black] (phi_gamma) circle (2pt);
		%
		\node [left = 6.5 of X.center] {$\color[HTML]{f0e442}\mathcal{X}$};
		\node [above left = -0.5 and 4 of G.center] {$\color[HTML]{8fd6c2}G$};
		%
		\node [above right = 0.4 and 4 of X.center] {$\color{light_blue}\mathcal{K}$};
		\node [below left = 1 and 2 of X.center] {\color{Mahogany}$G$-orbit of $x$};
		%
		\draw [->,thick] ([xshift=-4pt] x) to [bend right=45] node [midway, left] {$\varphi_g$} ([xshift=-3pt,yshift=3pt] phi);
		\draw [->,thick] ([xshift=-3pt,yshift=2pt] x) to [bend left=45] node [midway, left] {$\varphi_{\gamma}$} ([xshift=-5pt] phi_gamma);
		\draw [->,thick] ([xshift=5em,yshift=-1em] X.center) to [bend left=30] node [midway, right] {$\gamma$} ([xshift=0em,yshift=3.5em] G.center);
		\draw [->,thick] ([yshift=+3pt] gamma_phi) to [bend right=45] node [midway, right, rotate=-30, xshift=-10pt,yshift=8pt] {$(\cdot)\ast g$} ([xshift=+4pt] gamma);
	\end{tikzpicture}
	\caption{Equivariance of the moving frame~$\gamma$ and its construction via the cross-section~$\mathcal{K}$}
	\label{fig:moving_frame}
\end{figure}
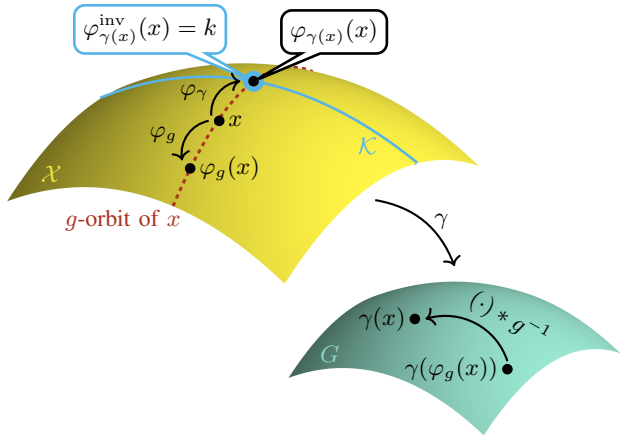

A \emph{moving frame} is a mapping $\gamma: \mathcal{X} \to G$ that has the following \emph{equivariance} property (illustrated in Fig.~\ref{fig:moving_frame}):
\begin{equation}\label{eq:equivariance}
	\gamma(\varphi_g(x)) \ast g = \gamma(x)
\end{equation}
It may be associated with a \emph{coordinate cross-section} $\mathcal{K}$ that transversely intersects $G$-orbits on $\mathcal{X}$. Informally, for an \mbox{$r$-dimensional} Lie group $G$ acting freely on the \mbox{$n$-dimensional} manifold $\mathcal{X}$, let  $\varphi_g^{\mathrm{inv}}$ be the part of $\varphi_g$ that maps to an \mbox{$r$-dimensional} submanifold of $\mathcal{X}$ such that it is invertible with respect to $g$ in a neighborhood of the identity element $e \in G$. Then, one can select a constant ${k}$ in the image of $\varphi_g^{\mathrm{inv}}$ that defines the unique point at which the $G$-orbit of a generic point $x$ intersects the $(n-r)$-dimensional cross-section $\mathcal{K}$. In other words, the moving frame can be obtained by solving the \emph{normalization equation}
\begin{equation*}
	\varphi_h^\mathrm{inv}(x) = k
\end{equation*}
for $h \in G$. The local solution $h = \gamma(x)$ defines the moving frame.


\subsubsection{Invariant Dynamics~\cite{bonnabelSymmetryPreservingObservers2008,boothbyIntroductionDifferentiableManifolds2003}}
Consider the dynamical control system 
\begin{equation}\label{eq:system_prelim}
	\dot{x} = f(x,u)
\end{equation}
where $x(t) \in \mathcal{X}$ (a differentiable manifold), $u(t) \in \mathcal{U}$ (a set), and $f(\cdot,u): \mathcal{X} \to \mathsf{T}_x \mathcal{X}$ (a \emph{vector field} on $\mathcal{X}$ for each $u \in \mathcal{U}$). Here, $\mathsf{T}_x \mathcal{X}$ denotes the tangent space to $\mathcal{X}$ at $x$. Let $\{ (\varphi_g(x),\psi_g(u)) \}_{g\in G}$ be a transformation group on $\mathcal{X} \times \mathcal{U}$, where $G$ is an $r$-dimensional Lie group. The mapping $\varphi_g: \mathcal{X} \to \mathcal{X}$ induces the \emph{tangent map} $\mathrm{T}_x \varphi_g: \mathsf{T}_x \mathcal{X} \to \mathsf{T}_{\varphi_g(x)} \mathcal{X}$ at $x$. Note that if $\mathcal{X} = \mathbb{R}^n$, then $\mathrm{T}_x \varphi_g$ is simply the Jacobian matrix, $\partial \varphi_g(x) / \partial x$. The system~\eqref{eq:system_prelim} is called \emph{$G$-invariant} if
\begin{equation*}
	f(\varphi_g(x),\psi_g(u)) = \mathrm{T}_x \varphi_g \big( f(x,u) \big)
\end{equation*}
It follows that the tangent map of $\varphi_{g \ast h}(x) = \varphi_g ( \varphi_h (x))$ satisfies
\begin{equation*}
	\mathrm{T}_x \varphi_{g \ast h} = \mathrm{T}_{\varphi_h (x)} \varphi_g \circ \mathrm{T}_x \varphi_h
\end{equation*}
A function $I: \mathcal{X} \times \mathcal{U} \to \mathbb{R}$ is called an \emph{invariant} if $I(\varphi_g(x),\psi_g(u)) = I(x,u)$ for all $g \in G$. Suppose $G$ acts freely on $\mathcal{X}$. Then, there locally exist $n-r$ functionally independent invariants $(I_1(x), \dots, I_{n-r}(x))$.

\subsection{Immersion and Invariance Observers~\cite{karagiannisNonlinearObserverDesign2005,astolfiChapter5Reducedorder2008}}
\label{sec:IandI}

Consider a dynamical system whose state is described by an unmeasured part, $x \in \mathcal{X} \subset \mathbb{R}^n$, and a measured part, $y\in\mathcal{Y}\subset\mathbb{R}^p$, with dynamics 
\begin{align*}
	\dot{x} &= f(x,y) \\
	\dot{y} &= h(x,y)
\end{align*}
The dynamical system
\begin{equation}\label{eq:alpha_prelim}
	\dot{z} = \alpha(z,y)
\end{equation}
where $z \in \mathbb{R}^{q(\geq n)}$, is called a (global) \emph{reduced-order observer} for $x$ if there exists a smooth manifold
\begin{equation}\label{eq:zero_error_prelim}
	\mathcal{Z} = \left\{ (x,y,z) ~|~ \theta(z,y) = \phi(x,y) \right\}
\end{equation}
defined by smooth mappings $\theta$ and $\phi$ that are left invertible with respect to their first argument such that $\mathcal{Z}$ is positively invariant and (globally) asymptotically attractive. The estimate of $x$ is then given by
\begin{equation}\label{eq:xhat_prelim}
	\hat{x} = \phi^{(\mathrm{L},\cdot)}(\theta(z,y),y)
\end{equation}
where $\phi^{(\mathrm{L},\cdot)}$ denotes the functional left inverse of ${\phi}$ with respect to its first argument; that is, $\phi^{(\mathrm{L},\cdot)}(\phi(x,y),y) = x$.

\section{INVARIANT PRE-OBSERVER}\label{sec:pre-observer}

Consider a system whose state is given by an unmeasured part $x \in \mathcal{X} \subseteq \mathbb{R}^n$ and a measured part $y \in \mathcal{Y}$. Here, $\mathcal{Y}$ is a $p$-dimensional differentiable manifold and $\mathcal{X}$ is an open subset of $\mathbb{R}^n$ that contains the origin. The dynamics of this system are given by
\begin{subequations}\label{eq:nonlinear_system}\begin{align}
	\dot{x} &= f(x,y,u) \\
	\dot{y} &= h(x,y,u)
\end{align}\end{subequations}
where $u\in\mathcal{U}$ is the known ``input'' to the system. It is not necessarily just composed of control inputs; rather it is a known signal on which a particular Lie group acts. Here, the dynamics of the measured part of the state, $y$, may be expressed intrinsically, that is, without specifying a local coordinate chart. 

We consider systems of the form~\eqref{eq:nonlinear_system} that are \emph{invariant} under the action of some Lie group $G$. 
\begin{assumption}\label{ass:transformation_group}
	The system~\eqref{eq:nonlinear_system} is $G$-invariant under the transformation group $\{(\varphi_g(x),\varrho_g(y),\psi_g(u))\}_{g\in G}$, where $G$ is an $r$-dimensional Lie group. That is,
	\begin{align*}
		\mathrm{T}_x \varphi_g \big( f(x,y,u) \big) &= f(\varphi_g(x),\varrho_g(y),\psi_g(u)) \\
		\mathrm{T}_y \varrho_g \big( h(x,y,u) \big) &= h(\varphi_g(x),\varrho_g(y),\psi_g(u))
	\end{align*}
	Furthermore, $\varphi_g(x)$ is linear in $x$.
\end{assumption}

We can now describe what it means for a reduced-order observer to be symmetry-preserving under the transformation group considered in Assumption~\ref{ass:transformation_group}. Briefly, a pre-observer is an observer for which there is not (yet) any claim about error convergence. We postulate a form for the observer of the unmeasured part of the state $x$ that preserves invariance of the state estimate dynamics. Inspired by~\cite{bonnabelSymmetryPreservingObservers2008}, we propose the following definition.
\begin{definition}[$G$-invariant reduced-order pre-observer]\label{def:pre-observer}
	The dynamical system
	\begin{equation}\label{eq:pre_observer}
		\dot{z} = \alpha(z,y,u)
	\end{equation}
	with output
	\begin{equation}\label{eq:xhat}
		\hat{x} = z + \beta(y)
	\end{equation}
	for some smooth map $\beta: \mathcal{Y} \to \mathcal{X}$ is a \emph{$G$-invariant reduced-order pre-observer} if the system
	\begin{equation}\label{eq:xhatdot}
		\dot{\hat{x}} = \alpha(\hat{x}-\beta(y),y,u) + \mathrm{T}_y \beta \big( h(x,y,u) \big)
	\end{equation}
	is $G$-invariant and the manifold
	\begin{equation}\label{eq:zero_error}
		\mathcal{Z} = \left\{ (z,x,y) \in \mathcal{X} \times \mathcal{X} \times \mathcal{Y} ~|~ z = x - \beta(y) \right\}
	\end{equation}
	is positively invariant. A $G$-invariant pre-observer is a \emph{$G$-invariant observer} if $\mathcal{Z}$ is asymptotically attractive.
\end{definition}

The prescription of the \emph{zero-error manifold} $\mathcal{Z}$ here is not quite as general as the case described in~\cite{astolfiChapter5Reducedorder2008}. We instead consider the condition that defines $\mathcal{Z}$ to be linear in $z$ and $x$ (and for $z$ to be the same dimension as $x$). In other words, we choose $\theta(z,y) = z$ and $\phi(x,y) = x - \beta(y)$ in \eqref{eq:zero_error_prelim}. This simplification reveals an intuitive choice for the \emph{observer map} $\beta$ in the following lemma based on the underlying geometry.
\begin{lemma}\label{lem:beta}
	Suppose there exists a moving frame $\gamma: \mathcal{Y} \to G$ that only depends on $y\in\mathcal{Y}$, and let $\ell: \mathcal{Y} \to \mathcal{X}$ be a smooth map. If
	\begin{equation}\label{eq:beta}
		\beta(y) = \varphi_{\gamma(y)^{-1}} \Big( \ell\big( \varrho_{\gamma(y)}(y) \big) \Big)
	\end{equation}
	then the following commutative identities (illustrated in Fig.~\ref{fig:beta}) hold for all $g\in G$ and $y\in\mathcal{Y}$:
	\begin{equation}
		\beta(\varrho_g(y)) = \varphi_g(\beta(y))
	\end{equation}
	\begin{equation}\label{eq:rho_tangent_map}
		\mathrm{T}_{\varrho_g(y)} \beta \circ \mathrm{T}_y \varrho_g = \mathrm{T}_{\beta(y)} \varphi_g \circ \mathrm{T}_y \beta
	\end{equation}
	In other words, $\beta$ commutes with the transformation group.
\end{lemma}
\begin{figure}[thpb!]
	\centering
	\begin{tikzpicture}
		\tikzstyle{every node}=[font=\small]
		%
		\node at (0,0) {\includegraphics[width=2.75in]{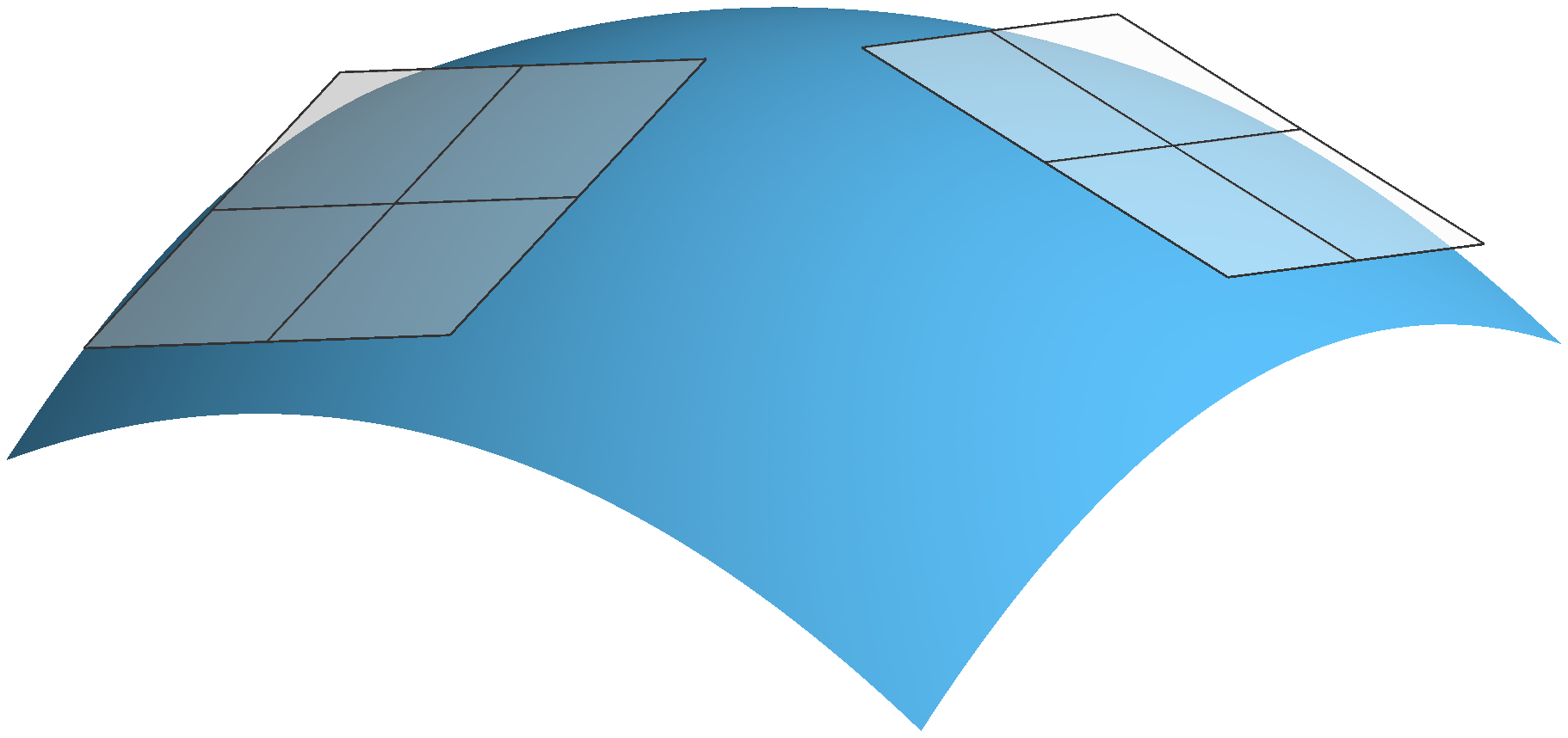}};
		\node at (0,-4) {\includegraphics[width=2.75in]{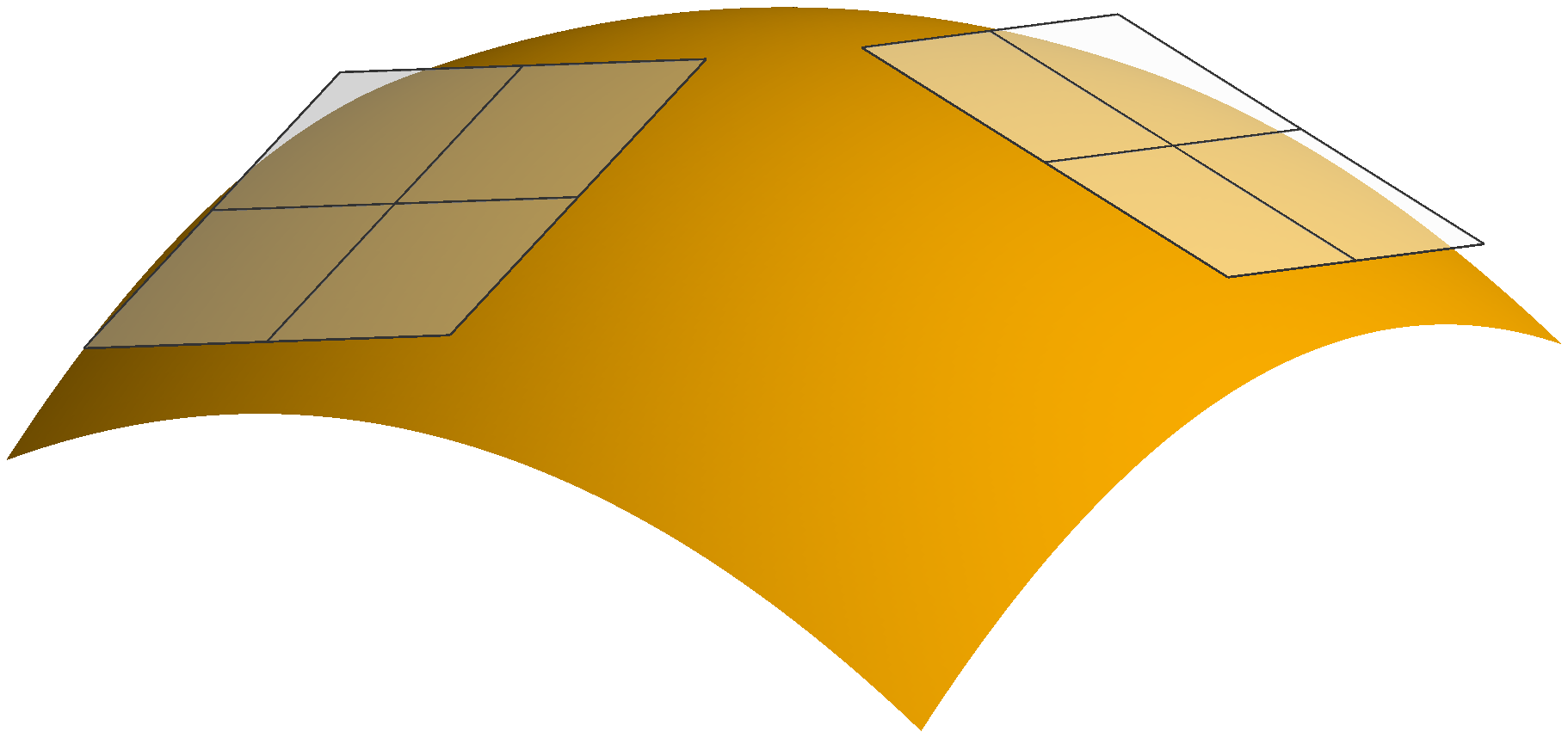}};
		%
		\coordinate[label={[xshift=-1ex,yshift=-0.2ex]$y$}] (y) at (-1.72,0.75);
		\coordinate[label={[xshift=2ex,yshift=0ex]$\varrho_g(y)$}] (rho) at (1.74,1);
		\coordinate[label={[left,xshift=0ex,yshift=-1ex]$\beta(y)$}] (beta) at (-1.72,-3.25);
		\coordinate (phi) at (1.74,-3);
		\node[rectangle callout,fill=white,opacity=0.8,text opacity=1,rounded corners,draw=Black,callout absolute pointer=(phi.south),below left= 3em and -3em of phi.south] {$\beta(\varrho_g(y)) = \varphi_g(\beta(y))$};
		\fill[black] (y) circle (2pt);
		\fill[black] (rho) circle (2pt);
		\fill[black] (beta) circle (2pt);
		\fill[black] (phi) circle (2pt);
		%
		\node at (0,-1.4) {$\color[HTML]{56b4e9}\mathcal{Y}$};
		\node at (0,-5.4) {$\color[HTML]{e69f00}\mathcal{X}$};
		%
		\coordinate (y_beta) at ([xshift=-0.5in,yshift=-0.5in] y);
		\coordinate (beta_y) at ([xshift=-0.5in] beta);
		\draw [->,thick,Mahogany] (y_beta) to [bend right=30] node [midway, left] {$\color{Black}\beta$} node [midway, right] {$\color{Black}\mathrm{T}_y \beta$} (beta_y);
		\coordinate (rho_phi) at ([xshift=+0.4in,yshift=-0.4in] rho);
		\coordinate (phi_rho) at ([xshift=+0.3in,yshift=0.1in] phi);
		\draw [->,thick,Mahogany] (rho_phi) to [bend left=30] node [midway, left] {$\color{Black}\beta$} node [midway, right] {$\color{Black}\mathrm{T}_{\varrho_g(y)} \beta$} (phi_rho);
		\coordinate (y_rho) at ([xshift=+4pt,yshift=+1pt] y);
		\coordinate (rho_y) at ([xshift=-4pt,yshift=+1pt] rho);
		\draw [->,thick,Mahogany] (y_rho) to [bend left=20] node [midway, above=-2pt] {$\color{Black}\varrho_g$} node [midway, below] {$\color{Black}\mathrm{T}_y \varrho_g$} (rho_y);
		\coordinate (beta_phi) at ([xshift=+4pt,yshift=-1pt] beta);
		\coordinate (phi_beta) at ([xshift=-4pt,yshift=-1pt] phi);
		\draw [->,thick,Mahogany] (beta_phi) to [bend right=15] node [midway, above] {$\color{Black}\varphi_g$} node [midway, below] {$\color{Black}\mathrm{T}_{\beta(y)} \varphi_g$} (phi_beta);
	\end{tikzpicture}
	\caption{Commutative relationship between $\beta$ and the transformation group}
	\label{fig:beta}
\end{figure}
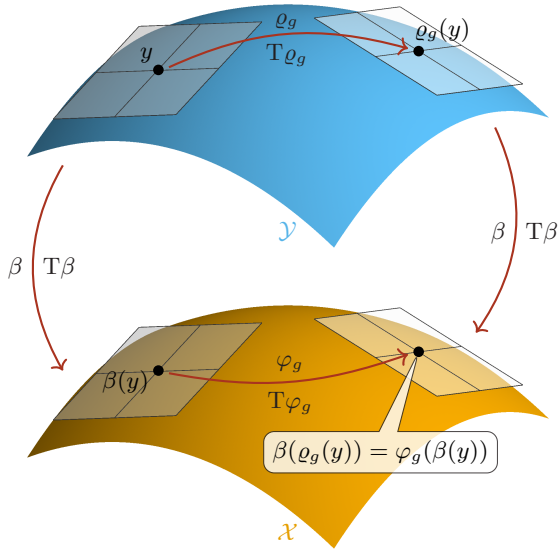 
\begin{proof}
	Beginning with the definition of $\beta$, we have
	\begin{equation*}
		\beta(\varrho_g(y)) = \varphi_{\gamma(\varrho_g(y))^{-1}} \Big( \ell\big( \varrho_{\gamma(\varrho_g(y))}(\varrho_g(y)) \big) \Big)
	\end{equation*}
	By the equivariance property~\eqref{eq:equivariance} of the moving frame,
	\begin{equation*}
		\beta(\varrho_g(y)) = \varphi_{(\gamma \ast g^{-1})^{-1}} \Big( \ell\big( \varrho_{\gamma \ast g^{-1}}(\varrho_g(y)) \big) \Big)
	\end{equation*}
	Using the composition properties for group elements and group actions,
	\begin{align*}
		\beta(\varrho_g(y)) &= \varphi_{g \ast \gamma^{-1}} \Big( \ell\big( \varrho_{\gamma \ast g^{-1} \ast g}(y) \big) \Big) \\
		&= \varphi_g \Big( \varphi_{\gamma^{-1}} \big( \ell\big( \varrho_\gamma(y) \big) \big) \Big) \\
		&= \varphi_g(\beta(y))
	\end{align*}
	Finally, \eqref{eq:rho_tangent_map} follows directly from the properties of the tangent map.
\end{proof}

Using Lemma~\ref{lem:beta}, a $G$-invariant pre-observer is readily constructed for a system satisfying Assumption~\ref{ass:transformation_group}. 
\begin{theorem}\label{thm:pre_observer}
	Suppose Assumption~\ref{ass:transformation_group} and the conditions of Lemma~\ref{lem:beta} hold. Let the vector field $\alpha(\cdot,y,u): \mathcal{X} \to \mathsf{T}\mathcal{X}$ be defined by
	\begin{equation}\label{eq:alpha}
		\alpha(z,y,u) = f(z+\beta(y),y,u) - \mathrm{T}_y \beta \big( h(z+\beta(y),y,u) \big)
	\end{equation}
	and let the observer map $\beta$ be given by~\eqref{eq:beta}. Then, the dynamical system~\eqref{eq:pre_observer} with output~\eqref{eq:xhat} is a $G$-invariant, reduced-order pre-observer.
\end{theorem}
\begin{proof}
	We begin by showing invariance of the state estimate dynamics~\eqref{eq:xhatdot}. Define	
	\begin{equation*}
		F(\hat{x},x,y,u) = \alpha(\hat{x}-\beta(y),y,u) + \mathrm{T}_y \beta \big( h(x,y,u) \big)
	\end{equation*}
	Then,
	\begin{multline*}
		\mathrm{T}_{\hat{x}} \varphi_g \big( F(\hat{x},x,y,u) \big) = \mathrm{T}_{\hat{x}} \varphi_g \big( f(\hat{x},y,u) \big) \\
		- \big( \mathrm{T}_{\hat{x}} \varphi_g \circ \mathrm{T}_y \beta \big) \big(  h(\hat{x},y,u) - h(x,y,u) \big)
	\end{multline*}
	The assumed linearity of $\varphi_g$ implies $\mathrm{T}_x \varphi_g$ does not depend on the choice of base point $x$. Therefore, $\mathrm{T}_{\hat{x}} \varphi_g = \mathrm{T}_{\beta(y)} \varphi_g$, and Lemma~\ref{lem:beta} can be used along with the invariance of $f$ to obtain
	\begin{multline*}
		\mathrm{T}_{\hat{x}} \varphi_g \big( F(\hat{x},x,y,u) \big) = f(\varphi_g(\hat{x}),\varrho_g(y),\psi_g(u)) \\
		- \big( \mathrm{T}_{\varrho_g(y)} \beta \circ \mathrm{T}_y \varrho_g \big) \big( h(\hat{x},y,u) - h(x,y,u) \big)
	\end{multline*}
	Since $h$ is also $G$-invariant, we have
	\begin{multline*}
		\mathrm{T}_{\hat{x}} \varphi_g \big( F(\hat{x},x,y,u) \big) = f(\varphi_g(\hat{x}),\varrho_g(y),\psi_g(u)) \\
		- \mathrm{T}_{\varrho_g(y)} \beta \big( h(\varphi_g(\hat{x}),\varrho_g(y),\psi_g(u)) \\
		- h(\varphi_g(x),\varrho_g(y),\psi_g(u)) \big)
	\end{multline*}
	By Lemma~\ref{lem:beta}, we recognize
	\begin{equation*}
		\varphi_g(\hat{x}) = \varphi_g(z) + \beta(\varrho_g(y))
	\end{equation*}
	Then, it follows that 
	\begin{multline*}
		\mathrm{T}_{\hat{x}} \varphi_g \big( F(\hat{x},x,y,u) \big) = \alpha(\varphi_g(\hat{x})-\beta(\varrho_g(y)),\varrho_g(y),\psi_g(u)) \\
		+ \mathrm{T}_{\varrho_g(y)} \beta \big( h(\varphi_g(x),\varrho_g(y),\psi_g(u)) \big)
	\end{multline*}
	Therefore,
	\begin{equation*}
		\mathrm{T}_{\hat{x}} \varphi_g \big( F(\hat{x},x,y,u) \big) = F(\varphi_g(\hat{x}),\varphi_g(x),\varrho_g(y),\psi_g(u))
	\end{equation*}
	That is, the system~\eqref{eq:xhatdot} is $G$-invariant. Next, we show the zero error manifold $\mathcal{Z}$ given in~\eqref{eq:zero_error} is positively invariant. Since $z - x + \beta(y) = 0$ on $\mathcal{Z}$, it is sufficient to verify that
	\begin{align*}
		&\alpha(x-\beta(y),y,u) - f(x,y,u) +  \mathrm{T}_y \beta \big( h(x,y,u) \big) \\
		&\quad = f(x,y,u) - \mathrm{T}_y \beta \big( h(x,y,u) \big) \\
		&\qquad - f(x,y,u) + \mathrm{T}_y \beta \big( h(x,y,u) \big) = 0
	\end{align*}
	Thus, referring to~\eqref{eq:pre_observer}, trajectories originating in $\mathcal{Z}$ remain in $\mathcal{Z}$. It follows that~\eqref{eq:alpha}--\eqref{eq:xhat} is a $G$-invariant reduced-order pre-observer.
\end{proof}

As an improvement over the general reduced-order observer described in \S\ref{sec:IandI}, Theorem~\ref{thm:pre_observer} leverages symmetry to construct a reduced-order pre-observer. The functions $\theta$, $\phi$, and $\alpha$ in~\eqref{eq:alpha_prelim}--\eqref{eq:xhat_prelim} are formulated using the moving frame~$\gamma$. The equivariance of the moving frame is the key property that makes this pre-observer $G$-invariant.

\section{INVARIANT OBSERVER}\label{sec:observer}
We now aim to find sufficient conditions for the pre-observer in Theorem~\ref{thm:pre_observer} to be a $G$-invariant reduced-order \emph{observer}. That is, we seek conditions under which $\mathcal{Z}$ is asymptotically attractive. Practically, this aim is accomplished by choosing the map $\ell:\mathcal{Y}\to\mathcal{X}$ such that a stability claim can be made about the origin of a state estimation error system. Like~\cite{bonnabelSymmetryPreservingObservers2008}, we consider nonlinear error coordinates that are $G$-invariant. Specifically, let
\begin{equation}
	\eta(z,x,y) = \varphi_{\gamma(y)}(z) + \ell(\varrho_{\gamma(y)}(y)) - \varphi_{\gamma(y)}(x)
\end{equation}
be invariant coordinates that are non-zero if and only if $(z,x,y) \notin \mathcal{Z}$. Thus, $\eta \to {0}$ as $t \to \infty$ implies $\mathcal{Z}$ is asymptotically attractive. Let $X = \varphi_{\gamma(y)}(x)$, $Y = \varrho_{\gamma(y)}(y)$, and $U = \psi_{\gamma(y)}(u)$. Using the moving frame to define these transformed points means $(X,Y,U)$ constitutes a complete set of invariants~\cite[Ch.~8]{olverClassicalInvariantTheory1999}. As will be shown shortly, the stability of the pre-observer~\eqref{eq:pre_observer} depends only $\eta$ and the invariants $X$, $Y$, and $Z$ (see Remark~\ref{rem:invariants}).

\begin{figure}[thpb!]
	\centering
	\begin{tikzpicture}
		\tikzstyle{every node}=[font=\small]
		%
		\node at (0,0) {\includegraphics[width=2.5in]{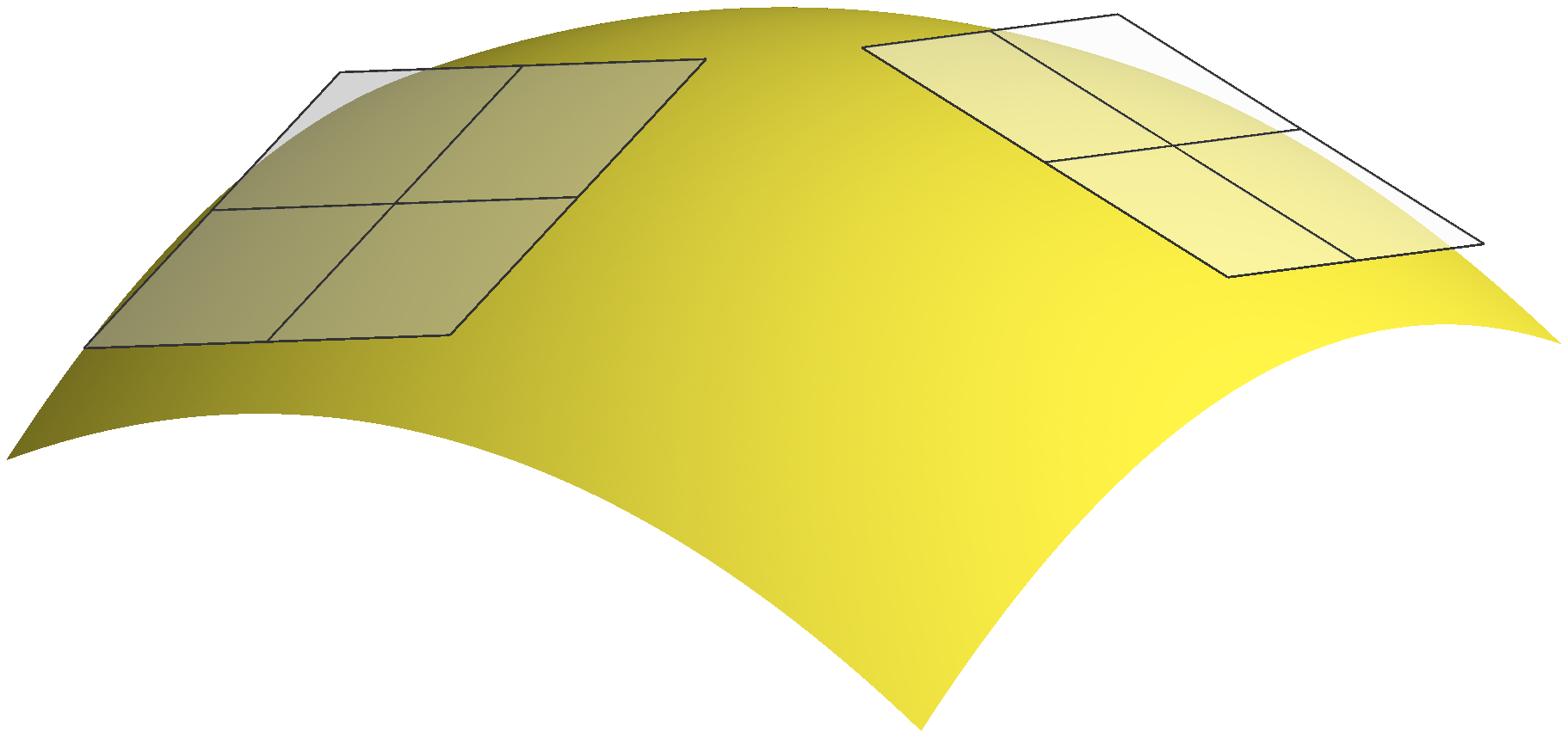}};
		\node at (0,-3) {\includegraphics[width=2in]{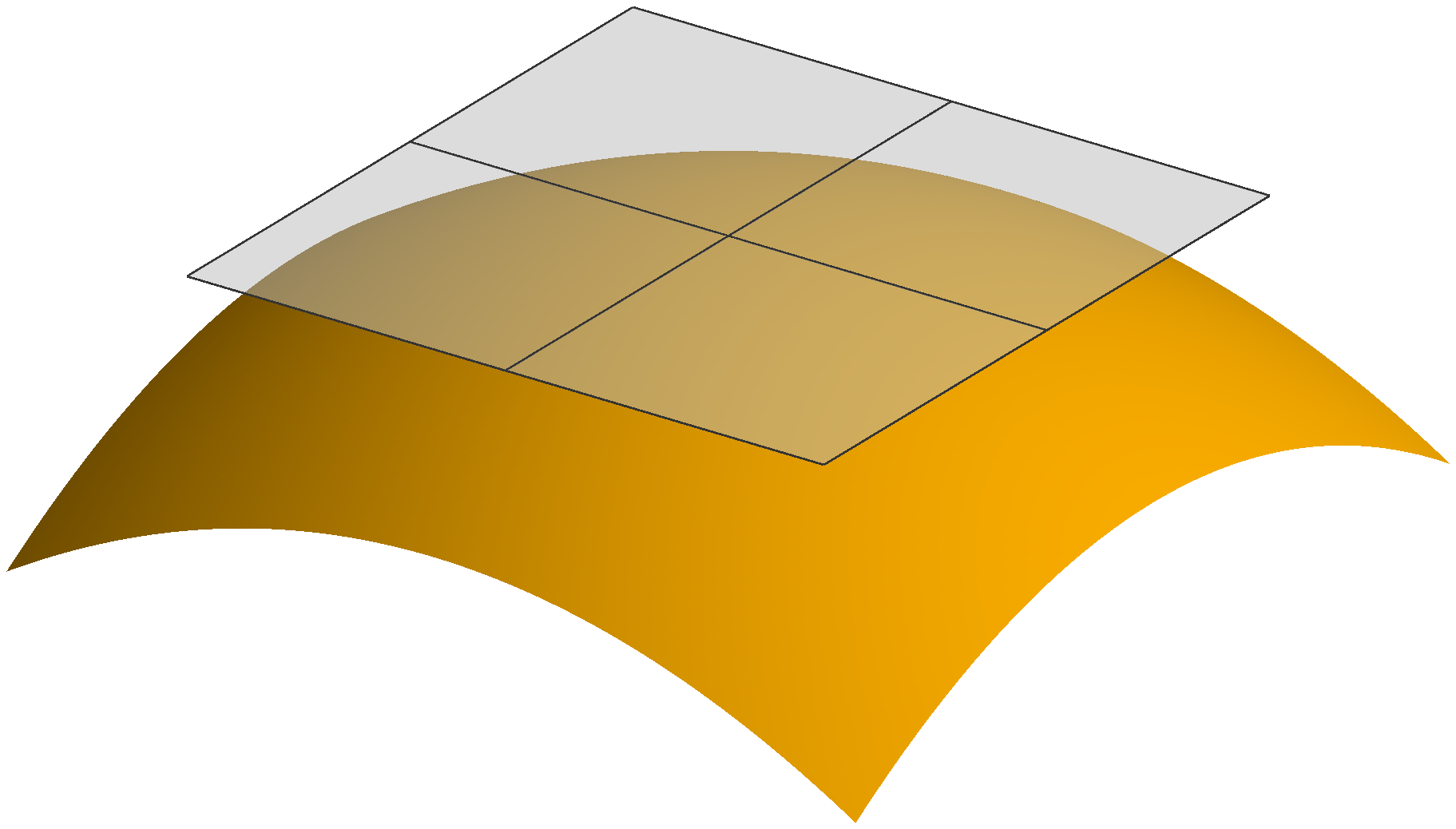}};
		%
		%
		\coordinate (y) at (-1.6,0.68);
		\node[rectangle callout,fill=white,opacity=0.8,text opacity=1,rounded corners,draw=Black,callout absolute pointer=(y.west),above left=1em and 0.2em of y.west] {$(y,\varphi_{g^{-1}}(\zeta))$};
		\fill[black] (y) circle (2pt);
		%
		\coordinate (x) at (-0.02,-2.37);
		\node[rectangle callout,fill=white,opacity=0.8,text opacity=1,rounded corners,draw=Black,callout absolute pointer=(x.west),below left=1.5em and -3em of x.west] {$\varphi_{\gamma(y)}(\varphi_{g^{-1}}(\zeta)) = \varphi_{\gamma(\varrho_g(y))}(\zeta)$};
		\fill[black] (x) circle (2pt);
		%
		\coordinate (rho_g) at (1.6,0.9);
		\node[rectangle callout,fill=white,opacity=0.8,text opacity=1,rounded corners,draw=Black,callout absolute pointer=(rho_g.east),below right=1em and 2em of rho_g.east] {$(\varrho_g(y),\zeta)$};
		\fill[black] (rho_g) circle (2pt);
		%
		\node at (0.45,-0.75) {$\mathcal{Y}(\times\mathcal{X})$};
		\node at (0.4,-4) {$\mathcal{X}$};
		\node at (-3,0.3) {$\color{gray}\mathsf{T}_y\mathcal{Y}$};
		\node at (2.7,1) {$\color{gray}\mathsf{T}_{\varrho_g(y)}\mathcal{Y}$};
		%
		\coordinate (y_x) at ([xshift=-1pt,yshift=-4pt] y);
		\coordinate (x_y) at ([xshift=-4pt] x);
		\draw [->,thick,Mahogany] (y_x) to [bend right=30] node [midway, left] {$\color{Black}\lambda(\cdot\,;\varphi_{g^{-1}}(\zeta))$} (x_y);
		\coordinate (rho_x) at ([xshift=1pt,yshift=-4pt] rho_g);
		\coordinate (x_rho) at ([xshift=4pt] x);
		\draw [->,thick,Mahogany] (rho_x) to [bend left=30] node [midway, right] {$\color{Black}\lambda(\cdot\,;\zeta)$} (x_rho);
		\coordinate (y_rho) at ([xshift=+4pt,yshift=+1pt] y);
		\coordinate (rho_y) at ([xshift=-4pt,yshift=+1pt] rho_g);
		\draw [->,thick,Mahogany] (y_rho) to [bend left=20] node [midway, below] {$\color{Black}(\varrho_g,\varphi_g)$} (rho_y);
		%
		\draw [-Stealth,very thick,NavyBlue] (y) --+ (0.8,0.5) node [above=1pt,fill=white,opacity=0.8,text opacity=1,draw=NavyBlue,rounded corners] {\color{black}$h(x,y,u)$};
		\draw [-Stealth,very thick,NavyBlue] (rho_g) --+ (-0.4,0.5) node [above right=1pt and -2em,fill=white,opacity=0.8,text opacity=1,draw=NavyBlue,rounded corners] {\color{black}$h(\varphi_g(x),\varrho_g(y),\psi_g(u))$};
		\draw [-Stealth,very thick,NavyBlue] (x) --+ (-0.3,0.7) node [right=7pt,fill=white,opacity=0.8,text opacity=1,draw=NavyBlue,rounded corners] {\color{black}Equation~\eqref{eq:lambda}};
	\end{tikzpicture}
	\vspace{-1em}
	\caption{Invariance of $\lambda$ and its tangent map}
	\label{fig:lambda}
\end{figure}
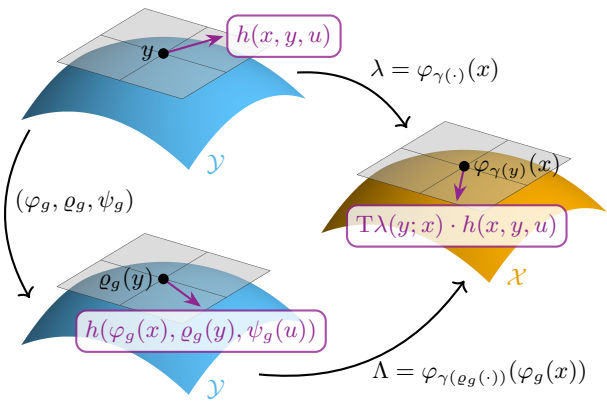

To derive sufficient conditions for asymptotic stability, we will make use of the following result.
\begin{lemma}\label{lem:lambda}
	Let $\lambda: \mathcal{Y} \to \mathcal{X}$ be the map
	\begin{equation*}
		\lambda(y;\xi) = \varphi_{\gamma(y)}(\xi)
	\end{equation*}
	where $\xi\in\mathcal{X}$ is held constant. Then,
	\begin{multline}\label{eq:lambda}
		\mathrm{T}_{(y;\varphi_{g^{-1}}(\zeta))} \lambda \big( h(x,y,u) \big) \\
		= \mathrm{T}_{(\varrho_g(y);\zeta)} \lambda \big( h(\varphi_g(x),\varrho_g(y),\psi_g(u)) \big)
	\end{multline}
	for any $g \in G$ and $\zeta\in\mathcal{X}$.
\end{lemma}
The following proof of Lemma~\ref{lem:lambda} is illustrated in Fig.~\ref{fig:lambda}.

\begin{proof}
	First, we recognize $\lambda$ is invariant since
	\begin{align*}
		\lambda(\varrho_g(y);\varphi_g(x)) &= \varphi_{\gamma(\varrho_g(y))}(\varphi_g(x)) \\
		&= \varphi_{\gamma(y) \ast g^{-1}}(\varphi_g(x)) \\
		&= \varphi_{\gamma(y)}(x) \\
		&= \lambda(y;x)
	\end{align*}
	for any $x\in\mathcal{X}$, where we again use the equivariance of the moving frame $\gamma$. Thus,
	\begin{equation*}
		\lambda(y;\varphi_{g^{-1}}(\zeta)) = \lambda(\varrho_g(y);\zeta)
	\end{equation*}
	for any $\zeta \in \mathcal{X}$. Since $\lambda$ is a composition of maps, it follows that for any $\zeta\in\mathcal{X}$,
	\begin{equation*}
		\mathrm{T}_{(y;\varphi_{g^{-1}}(\zeta))} \lambda = \mathrm{T}_{(\varrho_g(y);\zeta)} \lambda \circ \mathrm{T}_y \varrho_g
	\end{equation*}
	Applying this tangent map to the $G$-invariant vector field $h$, we obtain \eqref{eq:lambda}.
\end{proof}

Finally, sufficient conditions for~\eqref{eq:pre_observer} to be a $G$-invariant reduced-order observer are given as follows.
\begin{theorem}\label{thm:observer}
	Suppose the assumptions of Theorem~\ref{thm:pre_observer} hold. The $G$-invariant pre-observer~\eqref{eq:pre_observer} is a $G$-invariant observer if the origin $\eta = 0$ of the \emph{invariant error system}
	\begin{multline}\label{eq:error_system}
		\dot{\eta} = 
		f(X+\eta,Y,U) - f(X,Y,U) \\
		- \mathrm{T}_Y \beta \big( h(X+\eta,Y,U) - h(X,Y,U) \big) \\
		+ \mathrm{T}_{(Y;\eta)} \lambda \big( h(X,Y,U) \big)
	\end{multline}
	is asymptotically stable uniformly in $X$, $Y$, and $U$.
\end{theorem}
\begin{proof}
	By definition, the pre-observer~\eqref{eq:pre_observer} is an observer if the zero error manifold $\mathcal{Z}$ is positively invariant and asymptotically attractive or, equivalently, if the state estimation error dynamics have a globally asymptotically stable equilibrium at the origin. It remains for us to show that the estimation error dynamics are given by the invariant error system~\eqref{eq:error_system}. Since $\varphi_g(x)$ is linear in $x$, we can write 
	\begin{equation*}
		\eta = \varphi_{\gamma(y)}(z + \beta(y)) - \varphi_{\gamma(y)}(x)
	\end{equation*}
	Thus, the time derivative of $\eta$ satisfies
	\begin{multline*}
		\dot{\eta} = \mathrm{T}_{z + \beta(y)} \varphi_\gamma \Big( \alpha(z,y,u) + \mathrm{T}_y \beta \big( h(x,y,u) \big) \Big) \\
		- \mathrm{T}_x \varphi_\gamma \big( f(x,y,u) \big) + \mathrm{T}_{(y;z + \beta(y) - x)} \lambda \big( h(x,y,u) \big)
	\end{multline*}
	Substituting the definition of $\alpha$ from Theorem~\ref{thm:pre_observer} and again using the linearity of $\varphi_g(\cdot)$, we have
	\begin{multline*}
		\dot{\eta} = \mathrm{T}_{z + \beta(y)} \varphi_\gamma \big( f(z+\beta(y),y,u) \big) - \mathrm{T}_x \varphi_\gamma \big( f(x,y,u) \big) \\
		- \big( \mathrm{T}_{\beta(y)} \varphi_\gamma \circ \mathrm{T}_y \beta \big) \big( h(z+\beta(y),y,u) - h(x,y,u) \big) \\
		+ \mathrm{T}_{(y;{z} + \beta(y) - x)} \lambda \big( h(x,y,u) \big)
	\end{multline*}
	Applying the invariance of $f$ and $h$ through the use of Lemma~\ref{lem:beta} yields
	\begin{multline*}
		\dot{\eta} = f\big(\varphi_{\gamma}(z + \beta(y)),Y,U\big) -  f\big(X,Y,U\big) \\
		- \mathrm{T}_Y \beta \Big( h\big(\varphi_{\gamma}(z + \beta(y)),Y,U\big) -  h\big(X,Y,U\big) \Big) \\
		+ \mathrm{T}_{(y;z + \beta(y) - x)} \lambda \big( h(x,y,u) \big)
	\end{multline*}
	Notice the last term in the above equation can also be written as
	\begin{equation*}
		\mathrm{T}_{(y;\varphi_{\gamma^{-1}(y)}(\eta))} \lambda \big( h(x,y,u) \big)
	\end{equation*}
	Therefore, we can use Lemma~\ref{lem:lambda} along with a substitution of
	\begin{equation*}
		z = \varphi_{\gamma^{-1}}(\eta) - \beta(y) + x
	\end{equation*}
	to obtain \eqref{eq:error_system}.
\end{proof}

Theorem~\ref{thm:observer} states sufficient conditions for the reduced-order pre-observer constructed in Theorem~\ref{thm:pre_observer} to be an asymptotically stable observer. In particular, the error system~\eqref{eq:error_system} is $G$-invariant, meaning stability can be equivalently analyzed under arbitrary transformation by the Lie group's action.

\begin{remark}\label{rem:invariants}
	The error system~\eqref{eq:error_system} depends only on the invariant error $\eta$ and the invariants $X$, $Y$, and $U$, which can be reduced to a set of $n+p-r$ functionally independent invariants, $I(x,y,u)$~\cite[Ch.~8]{olverClassicalInvariantTheory1999}. This observation is consistent with the full-order case considered in~\cite[Theorem~3]{bonnabelSymmetryPreservingObservers2008}.
\end{remark}

\section{EXAMPLE: RIGID-BODY VELOCITY OBSERVER}\label{sec:example}

As an example, consider a rigid aircraft instrumented with an accelerometer, gyroscope, magnetometer, and GNSS receiver such that its position, $q$, and attitude rotation matrix, $R_\mathrm{IB}$, are known with negligible error. Furthermore, assume the angular velocity, $\omega$, and body-frame specific force, $a$, (obtained from filtered accelerometer readings) are available as inputs for the observer design. However, suppose that the body velocity, $v = (\mathrm{u,v,w})$ is not directly measured. The aim is to design a reduced-order velocity observer for the system
\begin{equation}\label{eq:rigid_body}
\begin{gathered}
	\underbrace{ \dot{v} }_{\dot{x}} = \underbrace{ v \times \omega + R_\mathrm{IB}\tp g + a }_{f(x,y,u)} \\
	\underbrace{ \begin{pmatrix}
		\dot{q} \\ \dot{R}_\mathrm{IB}
	\end{pmatrix} }_{\dot{y}} = \underbrace{ \begin{pmatrix}
		R_\mathrm{IB} {v} \\ R_\mathrm{IB} \cpem{{\omega}}
	\end{pmatrix} }_{h(x,y,u)}
\end{gathered}\end{equation}
where $\cpem{\cdot}$ is the skew-symmetric cross product equivalent matrix satisfying $\cpem{a}b = a \times b$ for 3-vectors $a$ and $b$ and $g$ is the gravity vector. 
\begin{proposition}
	The system~\eqref{eq:rigid_body} is $\mathsf{SO}(3)$-invariant with respect to the transformation group
	\begin{equation*}
		\varphi_g(x) = R_g v ,\quad \varrho_g(y) = \begin{pmatrix}
			q \\ R_\mathrm{IB} R_g\tp
		\end{pmatrix} ,\quad \psi_g(u) = \begin{pmatrix}
			R_g \omega \\ R_g a
		\end{pmatrix}
	\end{equation*}
	where $R_g \in G = \mathsf{SO}(3)$.
\end{proposition}
\begin{proof}
	We have
	\begin{align*}
		\mathrm{T}_x \varphi_g \big( f(x,y,u) \big) &= R_g (v \times \omega) + R_g R_\mathrm{IB}\tp g + R_g a \\
		&= R_g {v} \times R_g {\omega} + ({R}_\mathrm{IB} R_g\tp)\tp {g} + R_g {a} \\
		&= f(\varphi_g(x),\varrho_g(y),\psi_g(u))
	\end{align*}
	and
	\begin{align*}
		\mathrm{T}_y \varrho_g \big( h(x,y,u) \big) &= \begin{pmatrix}
			R_\mathrm{IB} v \\
			R_\mathrm{IB} \cpem{\omega} R_g\tp
		\end{pmatrix} \\
		&= \begin{pmatrix}
			R_\mathrm{IB} R_g\tp R_g v \\
			R_\mathrm{IB} R_g\tp \cpem{R_g \omega}
		\end{pmatrix} \\
		&= h(\varphi_g(x),\varrho_g(y),\psi_g(u))
	\end{align*}
	Here, we have used the property that $\cpem{R \xi} = R \cpem{\xi} R\tp$ for any $R\in\mathsf{SO}(3)$ and $\xi\in\mathbb{R}^3$.
\end{proof}

Since $R_\mathrm{IB}$ is an element of the Lie group $G$, the moving frame is simply
\begin{equation*}
	\gamma(y) = R_\mathrm{IB}
\end{equation*}
Because the transformation group is also linear in the measured part of the state, we can choose $\ell$ to be
\begin{equation*}
	\ell(y) = L {q}
\end{equation*}
where $L\in\mathbb{R}^{3 \times 3}$ is a tuning parameter. Therefore, 
\begin{equation*}
	\beta(y) = R_\mathrm{IB}\tp L q
\end{equation*}
Applying Theorem~\ref{thm:pre_observer}, we have
\begin{multline*}
	\alpha(z,x,y) = \underbrace{ (z+R_\mathrm{IB}\tp L q) \times {\omega} + R_\mathrm{IB}\tp g + a }_{f(z+\beta(y),y,u)} \\
	\underbrace{ + \cpem{\omega}R_\mathrm{IB}\tp L q - R_\mathrm{IB}\tp L R_\mathrm{IB} (z+R_\mathrm{IB}\tp L q) }_{-\mathrm{T}_y\beta \big( h(z+\beta(y),y,u) \big)}
\end{multline*}
with the estimate of $v$ given by
\begin{equation*}
	\hat{v} = z + R_\mathrm{IB}\tp L q
\end{equation*}
The sufficient condition given in Theorem~\ref{thm:observer} reduces to the requirement that the system
\begin{equation*}
	\dot{\eta} = -L \eta
\end{equation*}
is asymptotically stable. Therefore, if $(-L)$ is Hurwitz, then the pre-observer $\dot{z} = \alpha(z,y,u)$ is a globally exponentially stable, reduced-order, $\mathsf{SO}(3)$-invariant observer.

\begin{figure}[h!]
	\centering
	\includegraphics[scale=0.94]{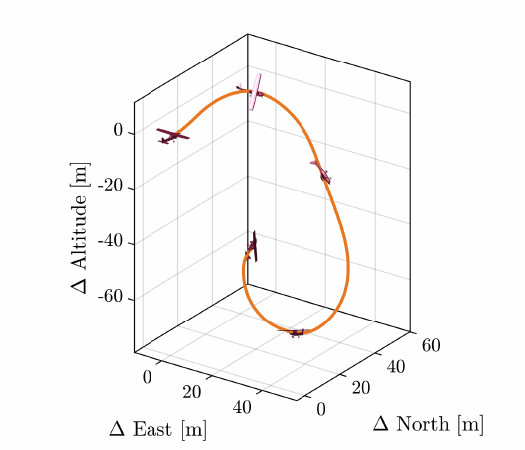}
	\caption{Maneuvering aircraft}
	\label{fig:maneuver}
\end{figure}

\begin{figure}[h!]
	\centering
	\vspace{1.1ex}
	\includegraphics[scale=0.94]{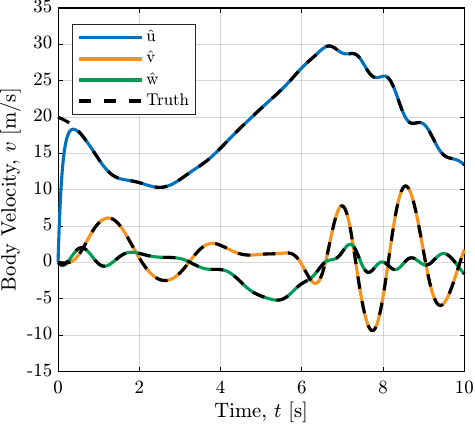}
	\caption{Velocity estimates}
	\label{fig:velocity_estimates}
\end{figure}

As a numerical demonstration, the aircraft equations of motion~\eqref{eq:rigid_body} were simulated using the nonlinear aerodynamic model in Simmons et~al.~\cite{simmonsAeropropulsiveModelingPropeller2023} to capture the specific force~$a$. Large-amplitude inputs to the aircraft's throttle, aileron, elevator, and rudder were commanded, resulting in maneuvering flight trajectory shown in Fig.~\ref{fig:maneuver}. 
Here, the aircraft starts in northbound, straight-and-level flight at 20~m/s with zero angle of attack; that is, $v(0)=[20~0~0]\tp$, $R_\mathrm{IB}(0)=\eye$, and $\omega(0)=0$. For the initial estimate $\hat{v}(0) = {0}$ and gain matrix $L = 10 \eye$, the time history of velocity estimates is shown in Fig.~\ref{fig:velocity_estimates}. 

\addtolength{\textheight}{-0ex}   

\begin{figure}[h!]
	\centering
	\includegraphics[scale=0.94]{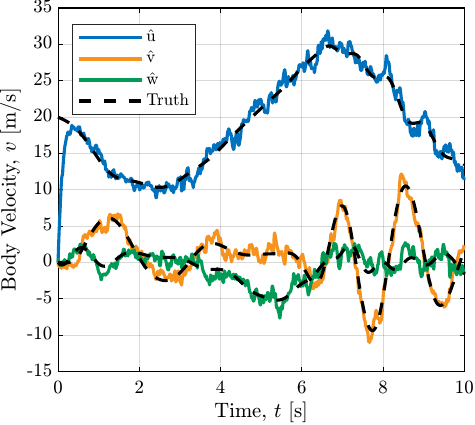}
	\caption{Velocity estimates with noisy measurements and inputs}
	\label{fig:velocity_estimates_noise}
\end{figure}

To stress the observer, we include noisy measurements of $y$ and $u$. Specifically, suppose 
\begin{align*}
	y_q &= q + w_q  &  u_\omega &= \omega + w_\omega \\
	y_{R_\mathrm{IB}} &= R_\mathrm{IB} \exp(\cpem{w_{R_\mathrm{IB}}})  &  u_a &= a + w_a
\end{align*}
where $w_q$, $w_{R_\mathrm{IB}}$, $w_\omega$, and $w_a$ are zero-mean, Gaussian, continuous-time, ``white noise'' with power spectral densities \mbox{$5\times10^{-4}\eye~\frac{\mathrm{m}^2}{\mathrm{Hz}}$}, \mbox{$10^{-7}\eye~\frac{1}{\mathrm{Hz}}$}, \mbox{$10^{-5}\eye~\frac{(\mathrm{rad}/\mathrm{s})^2}{\mathrm{Hz}}$}, and \mbox{$2\times10^{-2}\eye~\frac{(\mathrm{m}/\mathrm{s}^2)^2}{\mathrm{Hz}}$}, respectively. Figure~\ref{fig:velocity_estimates_noise} shows the velocity estimates when $y$ and $u$ are corrupted by a realization of these random processes, assuming a noise sampling rate of 1000~Hz to approximate continuous-time white noise.
Note that in this simple example, a nonlinear filter (such as an extended Kalman filter) produces less noisy estimates; reduced-order observers have a tendency to amplify measurement noise. Nonetheless, the results shown in Fig.~\ref{fig:velocity_estimates_noise} are indicative of the observer's inherent robustness to disturbances, as expected from the fact that the undisturbed invariant error system is globally exponentially stable~\cite[Lemma~5.1]{khalilNonlinearSystems1996}. As explored in~\cite{hopwoodNonlinearWindEstimation2025}, this local robustness also ensures that the disturbance incurred by using sampled measurements of $y$ yields a bounded result. In practice, ``measured'' values for low-rate signals, such as position data from GNSS, should be propagated between samples using an extended Kalman predictor, for instance. While proof of stability in the presence of noise is beyond the scope of this paper, the reader is directed to~\cite{hopwoodNoisetostateStableSymmetrypreserving2025} for an application-specific treatment illustrating how process noise might impact performance and stability analysis. 

\section{CONCLUSIONS}\label{sec:conclusions}

A symmetry-preserving, reduced-order observer has been presented. This approach is beneficial when part of the system's state is known with negligible error, avoiding unnecessary re-estimation of known signals and reducing computational complexity. Furthermore, the observer preserves symmetry. That is, the state estimate dynamics are invariant under the action of a Lie group. As a result, the state estimate error dynamics are also invariant under the group action. Exploiting these symmetries can simplify the selection of observer parameters as seen with the example of a rigid-body velocity observer. Tuning the proposed observer consists of finding a smooth map $\ell$ such that the origin of the invariant error system is asymptotically stable. For some systems, this problem is reduced to choosing a gain matrix, $L$, as shown in the example. By leveraging the geometry of the problem, the proposed observer simplifies both the design process and stability analysis, providing a powerful tool for safety-critical nonlinear systems.

\bibliographystyle{IEEEtran}
\bibliography{references}

\end{document}